\documentclass[11pt,twoside]{article}
\usepackage{amsmath}
 \usepackage{amssymb}
\usepackage{amsmath,amscd,graphics}
\usepackage[a4paper,textwidth=14cm,textheight=24cm,%
left=1.5in,right=1.5in,top=1in,bottom=1in,headsep=5mm]{geometry}

\newcommand{\bh}[1][i]{{(#1)}}
\newcommand{\eps}{\varepsilon}
\newcommand{\Real}{\mathbb{R}}
\newcommand{\ls}{\leqslant}
\newcommand{\norm}[1]{\left\|#1\right\|}

\newenvironment{proof}[1][Proof]{{\par \emph{#1.}}\;\;}{\hfill $\Box$\par}

 \def\be#1\ee{\begin{equation}#1\end{equation}}
 \def\bay#1\eay{\!\!\!\left\{\!\!\begin{array}{l}#1\displaystyle\end{array}\right.}
 \def\bln#1\eln{\begin{array}{l}#1\displaystyle\end{array}}

 \def\bma#1\ema{{\allowdisplaybreaks\begin{align}#1\end{align}}}
 \def\nnm{\notag}

 \def\bgr#1\egr{{\allowdisplaybreaks\begin{gather}#1\end{gather}}}
 \def\ef#1{(\ref{#1})}

       \newtheorem{lemma}{\bf Lemma}[section]
       \newtheorem{theorem}[lemma]{\bf Theorem}
       \newtheorem{proposition}[lemma]{\bf Proposition}

       \newtheorem{remark}[lemma]{\bf Remark}

\begin{document}
\abovedisplayskip=10pt plus 1pt minus 3pt \belowdisplayskip=10pt
plus 1pt minus 3pt 
\numberwithin{equation}{section} \allowdisplaybreaks

\renewcommand{\thefootnote}{\fnsymbol{footnote}}

\renewcommand{\thefootnote}{\fnsymbol{footnote}}
 \title{\Large\bf Long-time self-similar asymptotics of
  macroscopic quantum models }

 \title{\parbox{0.85\textwidth}{\Large\textbf{Long-time self-similar asymptotic of
the macroscopic quantum models}}\hspace*{0.15\textwidth}}
\author{\hspace*{.03\textwidth}
\begin{minipage}{.93\textwidth}{Hai-Liang Li,\!\!$^{1}$ \;
 Guo-Jing Zhang,\!\!$^2$\; Min Zhang,\!\!$^1$\; Chengchun
Hao$^3$}
\\[3mm]
\footnotesize\it $^{1}$Department of Mathematics, Capital Normal
University, Beijing 100037, P.R.China\\
 $^{2}$Department of Mathematics, Harbin Institute of
Technology,  Harbin 150001, P.R.China\\
$^3$Institute of Mathematics, Academy of Mathematics \& Systems
Science, CAS, Beijing 100190, P.R.China\\
 {\footnotesize\it   E-mail:~hailiang.li.math@gmail.com (H.L),
 zhanggj112@nenu.edu.cn (G.Z)
\\
\hspace*{3em} zhangminzi2004@163.com (M.Z), hcc@amss.ac.cn (C.H)}
 \end{minipage} }

\date{ }

\maketitle

\thispagestyle{empty} \pagestyle{myheadings} \markboth{H.-L. Li,
G.-J. Zhang, M. Zhang, C.C. Hao}{Long-time self-similar asymptotics}


\begin{abstract}
\noindent
The unipolar and bipolar macroscopic quantum models derived recently
for instance in the area of charge transport are considered in
spatial one-dimensional whole space in the present paper. These
models consist of nonlinear fourth-order parabolic equation for
unipolar case or coupled nonlinear fourth-order parabolic system for
bipolar case. We show for the first time the self-similarity
property of the macroscopic quantum models in large time. Namely, we
show that there exists a unique global strong solution with strictly
positive density to the initial value problem of the macroscopic
quantum models which tends to a self-similar wave (which is not the
exact solution of the models) in large time at an algebraic
time-decay rate.
\end{abstract}


\section {Introduction}
 \setcounter{equation}{0}

The quantum hydrodynamic (QHD) model for semiconductors is derived
and studied recently in the modelings and simulations of
semiconductor devices, where the effects of quantum mechanics
arise. The basic observation concerning the quantum hydrodynamics
is that the energy density consists of one additional new quantum
correction term of the order $O(\varepsilon)$ introduced first by
Wigner~\cite{Wigner32} in 1932, and that the stress tensor contains
also an additional quantum correction part~\cite{AncIaf89} 
related to the quantum Bohm potential~\cite{Bohm52}
 \be
 Q(\rho) =
-\frac{\varepsilon^2}{2m}\frac{\Delta\sqrt{\rho}}{\sqrt{\rho}},\label{disp-2}
 \ee
with observable $\rho>0$ the density, $m$ the mass, and
$\varepsilon$ the Planck constant. The quantum potential $Q$ is
responsible for producing the quantum behavior. Such possible
relation was also implied in the original idea initialized by
Madelung~\cite{Madelung27} to derive quantum fluid-type equations in
terms of Madelung's transformation applied to wave functions of the
Schr\"odinger equation of the pure state. Recently, the moment
method is employed to derive quantum hydrodynamic equations for
semiconductor device at nano-size based on the Wigner-Boltzmann (or
quantum Liouville) equation, refer to \cite{Jungel2001} for details.
For more important progress on the derivation of macroscopic quantum
models in terms of the entropy minimizer principle, one can refer to
the recent interesting works \cite{DR2003,Jungel2001,JM2005,JM2006}
and the references therein.

Starting with the quantum hydrodynamical models and performing the
relaxation limit asymptotical analysis, the macroscopic quantum
(Drift-Diffusion) model is derived rigorously~\cite{jlm06} for the
model of the unipolar carrier, the methods employed therein can be
generalized to general bipolar carriers. For positive charge
density, these models are indeed nonlinear fourth-order parabolic
equation for the unipolar case or the coupled nonlinear fourth-order
parabolic system for the bipolar case. We would like to mention that
according to the recent result in \cite{cl06}, the model~\ef{1.11}
can also be viewed as a relaxation limiting equation of the quantum
fluid model which can be derived by the nonlinear
Schr\"{o}dinger-Langevin equation, for which the rigorous short time
existence of weak solutions is proven recently in~\cite{JM485-495}.

We are interested in the long time asymptotical behavior of
solutions to the macroscopic quantum models in the present paper in
the one-dimensional real line, and we shall show that the global
classical solutions to the IVP~\eqref{b.1}--\eqref{b.1a} and the
IVP~\ef{1.11}--\ef{1.12} admit the character of self-similarity in
large time. In general, the typical bipolar macroscopic quantum
model widely used in semiconductor modeling in one dimension is the
following coupled nonlinear parabolic system
\def\bh{i}
\begin{align}
    &\left\{\begin{aligned}
     &\partial_t\rho_{\bh}-p(\rho_{\bh})_{xx}
     +\eps^2\left(
     \rho_{\bh}\left(
        \frac{(\sqrt{\rho_{\bh}})_{xx}}{\sqrt{\rho_{\bh}}}
        \right)_x\right)_x
     +(-1)^{i+1}(\rho_{\bh} E)_x=0,\quad   i=a,b,\\
     &E_x=\rho_{a}-\rho_{b},\qquad  t>0,\ x\in \Real,\
 \end{aligned}\right.    \label{b.1}
 \intertext{together with the initial data}
   &\quad \rho_{\bh}(x,0)=\rho_{\bh,_0}(x)>0,\quad x\in \Real,
   \quad \rho_{\bh,0}(\pm\infty)
     =\rho_{\pm}>0, \quad i=a,b,   \label{b.1a}
\end{align}
where $\rho_{a},\,\rho_b>0$ denote the macroscopic densities for
electron and hole respectively~\cite{Jungel2001}, $p(\rho_{\bh})$ is
the pressure function depending on the density $\rho_{\bh}$,
$\varepsilon>0$ is the scaled Planck constant, and $E$ denotes the
self-consistent electric field. We also use the symbols $(-1)^a=-1$
and $(-1)^b=1$ for the simplicity of statements.

In the absence of the electric filed (or in the so-called
quasi-neutral domain), the initial value (IVP) problem \ef{b.1}
reduces to the IVP problem for the  following unipolar macroscopic
quantum model
\begin{gather}
\label{1.11} \partial_t\rho-p_{xx}+\varepsilon^2
\mbox{$\left(\rho\left(
\frac{(\sqrt{\rho}\,)_{xx}}{\sqrt\rho}\right)_x\right)_x$}=0,\quad
t>0
\\
\label{1.12} \rho(x,0)=\rho_0(x)>0,\ x\in \Real,\quad
\rho_0(\pm\infty)=\rho_\pm>0,
\end{gather}
where $\rho=\rho(x,t)>0$ is the density of electron or hole and
$p=p(\rho)$ is the pressure function depending on $\rho$. It should
be noted that a similar model (the DDLS model), which takes the form
of \ef{1.11} but without the pressure term (i.e., $p=0$), also
arises in the study of interface fluctuations in spin systems, for
instance \cite{DLSS}.
\par

There are recently many analysis results on macroscopic quantum models of
the fourth-order parabolic type \ef{b.1} or \ef{1.11} and related models.
For the Eq.~\ef{1.11} without the density pressure function term
(the DLSS model~\cite{DLSS}), the positive classical solutions are
proven locally in-time in one-dimensional periodic domain
\cite{BL1994}, and the global existence of a spatially periodic $H^1$
solution and its exponential convergence to an equilibrium state is
shown \cite{MC2005} in terms of the entropy method and the
Csiszar-Kullback inequality for ``small" initial data. This is
mainly due to the failure of the maximum principle which makes it
impossible to establish a-priorily the upper and lower bounds of the
density and obtain the global in-time existence of solutions with the
strictly positive density. This, however, leads to the interesting
results on the global existence of a nonnegative weak solution, which
is first established in a one-dimensional bounded domain with
the Dirichlet and Neumann boundary condition~\cite{JP2000} where an
interesting entropy estimate is introduced to show the global
existence. Since then, some additional first order entropies are
also obtained \cite{JM2006,JV2007}.  More recently, the global
existence of multi-dimensional nonnegative weak solutions and their
exponential decay to an equilibrium state is also shown for the DLSS
model in a periodical domain~\cite{JM2008} based on the extended
multi-dimensional algorithmic entropy construction argument, and for
the DLSS model with an additional term of the given drift
potential~\cite{GST} in the framework of variation and Wasserstein's
metric subject to the finite initial mass. For more analysis related to
the DLSS model or Eq.~\ef{1.11} about numerical simulations or long
time convergences, one can refer to the recent papers
\cite{CJ2003,GJT2006,JM2008p,JT2003} and references therein. As for
the bipolar quantum model \ef{b.1}, the existence of a stationary
state is only analyzed recently \cite{U69-88}. Some interesting
quasi-neutral limit has been analyzed recently \cite{JV2007p}.
\par

However, there are few results on the global existence of classical
(strong) solutions with the strictly positive density and the long time
asymptotical behaviors of classical solutions for the macroscopic
quantum models \ef{b.1} and \ef{1.11} in the whole spatial space,
although there are a short time classical solution with the positive
density for the DLSS model~\cite{BL1994} and a global existence for
nonnegative weak solutions~\cite{JP2000,JM2006,JV2007,GST,JM2008p}.
The main difficulties in dealing with the macroscopic quantum models
\ef{b.1} and \ef{1.11} consist of the strong nonlinearity, the
degeneracy at vacuums, and the failure of the maximum principle, and the
coupling and interaction between the two carriers for the bipolar case.

It should be noted that it seems not obvious how to generalize the
framework of entropy estimates and/or Wasserstein's metric, used for
instance in \cite{GST,GJT2006,JM2006} to establish the global existence
of nonnegative solutions with the finite initial mass, to show the
global existence of a classical solution with the strictly positive
density for Eqs.~\ef{b.1} or \ef{1.11} in the whole spatial space
subject to the infinite initial mass (the case to be dealt with in the
present case), since in general Poincar\'e's inequality failed
and it is not trivial to establish the a-priori uniform control of the
density with respect to the time in order to understand the long time
behavior of global solutions.
\par

In this paper, we are interested in the large time asymptotical
behavior of the solution to the IVP
\eqref{b.1}--\eqref{b.1a} for the bipolar case and the IVP
\ef{1.11}--\ef{1.12}  for the unipolar  case, and we shall show that
the global classical solutions to IVP~\eqref{b.1}--\eqref{b.1a} and
IVP~\ef{1.11}--\ef{1.12} admit the character of self-similarity in
large time so long as it is around the self-similar wave initially.
To be more precise, let's introduce the quasi-linear parabolic
equation
\begin{eqnarray}
\label{1.14} \rho_t=p(\rho)_{xx},\ \ \ p'(\rho)>0.
\end{eqnarray}
It is well-known that Eq.\ef{1.14} has a unique self-similar
solution $W(x,t)$ up to a position shift (see\cite{ctlla})
\begin{eqnarray}
\label{1.15} \rho(x,t)=:W(\xi),&\xi=\frac{x}{\sqrt{t+1}},\quad
W(\pm\infty)=\rho_\pm,
\end{eqnarray}
and the solution $W(\xi)$ is increasing if $\rho_-<\rho_+$ and
decreasing if $\rho_->\rho_+$, and $W_{\xi}\rightarrow 0$ as
$|\xi|\rightarrow\infty$.
\par

To begin with, let us consider the case of regular initial data and
take the initial datum of the IVP~\eqref{b.1}--\eqref{b.1a}  and
IVP~\ef{1.11}--\ef{1.12} close to the self-similar solution $W$ in
some Sobolev norm. Then, we show in terms of the energy method below
that there exists a global unique classical solution to the
IVP~\eqref{b.1}--\eqref{b.1a} or the IVP~\ef{1.11}--\ef{1.12}, which
in particular tends to the self-similar wave $W$ in large time with an
algebraic decay rate.

We should also mention that it is not obvious so far how to
generalize the entropy functional or Wasserstein's metric technique as
used in \cite{GST,GJT2006,JM2006,JT2003} to prove the convergence of
global nonnegative solutions of the Eqs.~\ef{b.1} and \ef{1.11} to
the self-similar wave since the self-similar wave $W$ is not a
solution of the fourth-order quantum models \ef{b.1} and \ef{1.11}
and the convergence itself is a singular process in large time.
Moreover, some additional information on the lower and upper bounds of
the density are needed to prevent the possible appearance of the
singularity of the fourth order differential operator near vacuums.
This is nontrivial however due to the failure of the maximum principal
theory for the fourth-order equation. It is also interesting to
whether or not the global classical solution of the quantum models
\ef{b.1} and \ef{1.11} shall converge to the self-similar wave $W$
in large time for general initial data, instead of the small
perturbation of the self-similar wave, it is left for further
investigation.

We first investigate the long time asymptotical behavior of global
solutions to the IVP  for the unipolar equation
\ef{1.11}--\ef{1.12}, and then discuss the corresponding IVP
for the bipolar model \ef{b.1}--\ef{b.1a}.

Let
 \be
z_0(x)=\displaystyle\int_{-\infty}^x(\rho_0(y)-W(y+x_0))dy,\label{1.16}
 \ee then
$$
z_{0x}=\rho_0(x)-W(x+x_0),
$$
where $x_0$ is determined by
$$
\displaystyle\int_{-\infty}^{+\infty}(\rho_0(x)-W(x+x_0))dx=0.
$$

For the unipolar case, our main result on the global solution and
its large time behavior of IVP \ef{1.11}--\ef{1.12} is given as
follows.
\begin{theorem}\label{theorem 1.1}
 Let $p'(\rho)>0$ for $\rho>0$. Assume that
$\delta=:|\rho_+-\rho_-|\ll1$ and $z_0\in H^3(\Real)$ with
$\delta_0=:\|z_0(x)\|_{H^3(\Real)}$ small enough, 
then there is a  unique global
strong solution $\rho>0$ of IVP~\eqref{1.11}-\eqref{1.12} such that
$$
\rho-W\in L^{\infty}([0,\infty);H^2(\Real))\cap
L^2([0,\infty);H^4(\Real)),
$$
and the solution $\rho$ converges to the self-similar wave
$W(\frac{x+x_0}{\sqrt{t+1}})$ of the Eq.~\eqref{1.14} with an
algebraic time decay rate
\begin{gather*}
\|\partial_x^k(\rho-W)(t)\|_{L^2(\Real)}\ls C(1+t)^{-\frac{k+1}{2}},
 \quad \ \ k=0,1,2,
\\
\|(\rho-W)(t)\|_{L^{\infty}(\Real)}\ls C(1+t)^{-\frac{3}{4}},
\end{gather*}
where $C$ is a positive constant dependent of $\delta$ and
$\delta_0$.
\end{theorem}

Next, we state the main result on the convergence to the
self-similar wave  for the bipolar case. Although it leads to
additional difficulties, the coupling between carriers in \ef{b.1} may
cause some cancelation, and it is not clear that both the densities
of the bipolar QDD \ef{b.1} behave still or not like the unipolar
one for a small perturbation.

Denotes
 \be
z^{\bh}_0(x)=\displaystyle\int_{-\infty}^x(\rho_{\bh,0}(y)
-W(y+x_{0}))dy,
\quad i=a,b,
 \label{1.16a}
 \ee then
$$
z^{\bh}_{0x}=\rho_{\bh,0}(x)-W(x+x_{0}),\quad i=a,b,
$$
where $x_{0}$ is determined by
$$
\displaystyle\int_{-\infty}^{+\infty}(\rho_{\bh,0}(x)-W(x+x_{0}))dx=0.
$$
\begin{theorem}\label{theorem 1.2}
Let $p'(\rho)>0$ for $\rho>0$, and $\delta=:|\rho_+-\rho_-|\ll 1$.
Assume that $\inf_{x\in \Real}\rho_{\bh,0}>0$ $(i=a,b)$ with
$\int(\rho_{a,0}-\rho_{b,0})dx=0$, and $z_0^{\bh}\in H^3(\Real)$
with $\delta_0=:\|z^{a}_0\|_{H^3(\Real)}+\|z^{b}_0\|_{H^3(\Real)}$
small enough, then there is a unique global strong solution
$(\rho_{a}, \rho_{b},E)$ of the IVP~\eqref{b.1}-\eqref{b.1a} with
$\rho_a>0,\,\rho_b>0$ such that
$$
\rho_{\bh}-W\in L^{\infty}([0,\infty);H^2(\Real))\cap
L^2([0,\infty);H^4(\Real)), \quad  i=a,b,
$$
and both $\rho_{a}$ and $\rho_{b}$ converge to the self-similar wave
$W(\frac{x+x_0}{\sqrt{t+1}})$  of the Eq.~\eqref{1.14} with an
algebraic time decay rate
\begin{gather*}
  \|\partial_x^k(\rho_{a}-W)(t)\|_{L^2(\Real)}
  +\|\partial_x^k(\rho_{b}-W)(t)\|_{L^2(\Real)}\ls
  C(1+t)^{-\frac{k+1}{2}}, \quad
 \ \ k=0,1,2,
\\
  \|(\rho_{a}-W)(t)\|_{L^{\infty}(\Real)}
 +\|(\rho_{b}-W)(t)\|_{L^{\infty}(\Real)}\ls C(1+t)^{-\frac{3}{4}},
\\
\norm{E(t)}_{H^1(\Real)}\ls Ce^{-\beta t},
\end{gather*}
where $C>0$ and $\beta>0$ are constants dependent of $\delta$ and
$\delta_0$.
\end{theorem}
\vspace{2mm}

\textbf{Notations.\ }  $L^p(\Real)$ and $H^k(\Real)$ denote the
usual Lebesgue integrable functions space and the
 Sobolev space with norm $\|\cdot\|_{L^p(\Real)}$ and
$\|\cdot\|_{H^k(\Real)}$ respectively. we also use $\|\cdot\| $ to
denote $\|\cdot\|_{L^2(\Real)}$ for simplicity. $C$ and $c$ are used
to denote general positive constants.

\section{Proof of main results}

We shall prove Theorems~\ref{theorem 1.1}-\ref{theorem 1.2} in this
section. The key is to establish the a-priori estimates for short
time strong solutions. Without the loss of generality, we establish
the expected estimates in order to prove Theorem~\ref{theorem 1.1}
in Sect.~\ref{unipolar}, and show how to derive the estimates about
the electric filed in the proof of Theorem~\ref{theorem 1.2} in
Sect.~\ref{bipolar}.

\subsection{The unipolar case}
\label{unipolar}
In this section, we shall transform the primary equations in order
to study the existence and in particular its large time behavior of
the global solutions of the IVP~\ef{1.11}-\ef{1.12}.

Denote
\begin{equation}
\label{2.1}
z(x,t)=\int_{-\infty}^x\left(\rho(y,t)-W\left(\frac{y+x_0}{\sqrt{t+1}}\right)\right)dy,
\end{equation}
then
\begin{equation}
\label{2.2} z_x=\rho(x,t)-W(\frac{x+x_0}{\sqrt{t+1}}).
\end{equation}

We will derive the fourth order parabolic equation for $z$. Since
$W$ satisfies
\begin{equation}
\label{2.3} W_t=p(W)_{xx},
\end{equation}
and $\rho=\rho(x,t)$ satisfies Eq.\ef{1.11},
we have
\begin{equation}
\label{2.4} (\rho-W)_t-(p(\rho)-p(W))_{xx}
+\varepsilon^2\left(\rho\left(\frac{(\sqrt\rho)_{xx}}
{\sqrt\rho}\right)_x\right)_x=0.
\end{equation}

Integrating \ef{2.4} over $(-\infty,x)$ with respect to the spatial
variable and assuming $\rho_{xx}\to0$ as $|x|\to\infty$,  we obtain
from \ef{2.1} and \ef{2.2} the parabolic equation of the fourth
order for $z$ of the following form
\begin{equation}
\label{2.5}
z_t-(p'(W)z_x)_x+\frac{\varepsilon^2}{2}z_{xxxx}=(f_1+f_2)_x,
\end{equation}
with the initial datum
\begin{equation}
\label{2.6} z(x,0)=z_0(x),
\end{equation}
where
\bma
f_1&=\frac{\varepsilon^2}{2}\frac{(W_x+z_{xx})^2}{W+z_x}-\frac{\varepsilon^2}{2}W_{xx},
\label{2.7}\\
 f_2&=p(z_x+W)-p'(W)z_x-p(W)\label{2.8}.
\ema
Note that we have used the fact
$$\rho\left(\frac{(\sqrt\rho)_{xx}}{\sqrt\rho}\right)_x
=\frac{1}{2}\rho_{xxx}-
\frac{1}{2}\left(\frac{\rho_x^2}{\rho}\right)_x.$$

The existence of the global solution and the large time behavior for
the IVP \ef{2.5}--\ef{2.6} is obtained by the following proposition.

\begin{proposition}{\label{proposition 2.1}}
Let $p'(\rho)>0$ for $\rho>0$ and $\delta=|\rho_+-\rho_-|\ll1$.
Assume that $\|z_0\|_{H^3(\Real)} \ls \delta_0$ with $\delta_0>0$
sufficiently small, then there is a unique global strong solution
$z$ to the IVP~\eqref{2.5}-\eqref{2.6} satisfying
 \begin{gather*}
z\in L^{\infty}([0,\infty);H^3(\Real))\cap
L^2([0,\infty);H^5(\Real)),
\\
 \|\partial_x^kz(\cdot,t)\|_{L^2(\Real)} \ls
C(\delta,\delta_0)(1+t)^{-\frac{k}{2}},\ \ \ k=0,1,2,3,
\\
\|\partial_x^kz(\cdot,t)\|_{L^{\infty}(\Real)}\ls
C(\delta,\delta_0)(1+t)^{-\frac{1+2k}{4}},\ \ \ k=0,1,2,
\end{gather*}
where $C(\delta,\delta_0)>0$ is a positive constant depending only
on $\delta$ and $\delta_0$.
\end{proposition}

 \begin{remark}
 It is sufficient to prove Theorem~\ref{theorem 1.1}
 in terms of  Proposition~\ref{proposition 2.1}
  due to the relation between $\rho$ and $z$
  $$\rho=W+z_x$$
  and the transformation of \ef{2.2}--\ef{2.5}.
  The positivity of $\rho$ can be assured by the positivity
  of $W$ and the smallness of $z_x$. From the above, we also have
  $$\|\partial_x^k(\rho-W)(t)\|_{L^2(\Real)}
  =\|\partial_x^{k+1}z(t)\|_{L^2(\Real)},\ \ \ k=0,1,2. $$
\end{remark}

\bigskip

In order to prove Proposition \ref{proposition 2.1}, let us assume
that for the local in-time solution and $T>0$
\begin{equation}
\label{3.1} \delta_T=\max\limits_{k=0,1,2,3 }\sup\limits_{0\ls t\ls
T}(1+t)^{\frac{k}{2}}\|\partial_x^kz\|\ll1.
\end{equation}
By the Nirenberg's inequality and the above assumption, we have
 \be
 \|\partial_x^kz\|_{L^{\infty}(\Real)}
 \ls c\delta_{T}(1+t)^{-\frac{2k+1}{4}},\ \ \ k=0,1,2.\label{3.1a}
 \ee

The theorem for the existence of the local in-time solution is

\begin{theorem}{\label{theorem 3.1}}
Let $p'(\rho)>0$ for $\rho>0$ and assume that $\|z_0\|_{H^3(\Real)}$
small enough and $\inf\limits_{x\in \Real}({z_0}_x+W(x+x_0))>0$
$(i.e.\inf\limits_{x\in \Real}\rho_0>0)$. Then there exists a
$T^*>0$ such that there is a unique local solution of the
IVP~\eqref{2.5}-\eqref{2.6} for $t\in(0,T^*)$ satisfying
$$
\|z(\cdot,t)\|_{H^3(\Real)}<\infty,\ \
\rho(x,t)=z_x(x,t)+W(x+x_0,t)>0.
$$
\end{theorem}

  The proof of Theorem \ref{theorem 3.1} can be obtained by
 a standard method (see \cite{GST}), we omit it.

Now, we list the $L^p$-estimates of the derivatives of $W$ and
$n=\sqrt{W}$ as follows.
\begin{lemma}{\label{lemma 3.2}}
Let $W$ be the self-similar solution of \ef{1.14} and $n=\sqrt W$,
then it holds that (see\cite{cl06})
 \bgr \label{3.2} \|\partial_x^j
W(\cdot,t)\|_{L^p(\Real)}\ls
C\delta(1+t)^{-\frac{j}{2}+\frac{1}{2p}},
\\
 \label{3.3}
\|\partial_x^jn(\cdot,t)\|_{L^p(\Real)}\ls
C\delta(1+t)^{-\frac{j}{2}+\frac{1}{2p}},
 \egr
 for $j\geq
0,p\in[1,+\infty]$, $C>0$ is some constant.
\end{lemma}

For $f_1,f_2$, we have the following estimates.
\begin{lemma}{\label{lemma 3.3}}
Under the assumption \ef{3.1}, it holds for $f_1,f_2$
 \bma
&f_1=O(\delta_T+\delta)z_{xx}+O(\delta)r_2, \label{3.4}\\
&f_2=O(\delta_T)z_x,\label{3.5}
\ema
 where the function $r_k(x,t)$ is related to the
$k$th order derivative of $W$ with respect to $x$, which satisfies
by the definition
\begin{equation}\label{3.6}
 \|r_k(\cdot,t)\|_{L^p(\Real)}\ls
C(1+t)^{-\frac{k}{2}+\frac{1}{2p}},\ k=0,1,2,\cdots, \text{ and }
p\in[1,+\infty].
 \end{equation}
 \end{lemma}
\begin{proof} From Lemma \ref{lemma 3.2}, \ef{2.7} and \ef{2.8}, we have
 \begin{eqnarray*}
f_1&=&\frac{\varepsilon^2}{2}\frac{(W_x+z_{xx})^2}
{W+z_x}-\frac{\varepsilon^2}{2}
W_{xx}\\
&=&O(1)(W_x^2+2W_x\cdot z_{xx}+z_{xx}^2+W_{xx})\\
&=&(2W_x+z_{xx})\cdot z_{xx}+O(\delta)r_2\\
&=&O(\delta_T+\delta)\cdot z_{xx}+O(\delta)r_2, \\
f_2&=&p(z_x+W)-p'(W)z_x-p(W)\\
&=&\frac{p''(W)}{2}\cdot z_x^2+O(z_x^3)\\
&=&Cz_x(z_x+z_x^2)=O(\delta_T)z_x,
\end{eqnarray*}
by which the proof is easy.
\end{proof}

\begin{lemma}{\label{lemma 3.4}}
 Under the assumption \ef{3.1}, it holds for the
local in-time solution $z$
 \bma
\|z(t)\|^2+\displaystyle\int_0^t\|z_{x}(s)\|^2ds+ \int_0^t
\|z_{xx}(s)\|^2ds \ls O(\delta+\delta_0)^2,\label{3.7} \ema for
$0\ls t\ls T$, provided that $\delta_T+\delta$ is small enough.
  \end{lemma}
\begin{proof}
Taking the $L^2$-inner product of \ef{2.5} with $z$, we get with the
help of the integration by parts
\begin{eqnarray*}
\int_{\Real} [z_t-(p'(W)z_x)_x+\frac{\varepsilon^2}{2}z_{xxxx}]\cdot
zdx &=&\int_{\Real} z_t\cdot
z+p'(W)z_x^2+\frac{\varepsilon^2}{2}z_{xx}^2dx\\
&=&\frac{1}{2}\frac{d}{dt}\|z\|^2
+\frac{\varepsilon^2}{2}\|z_{xx}\|^2+\int_{\Real} p'(W)z_x^2dx.
\end{eqnarray*}
By Lemma \ref{lemma 3.3} for $f_1,f_2$ and Cauchy's inequality, we
get
\begin{eqnarray*}
\int_{\Real} (f_1+f_2)_x\cdot
zdx&=&\int_{\Real} -(f_1+f_2)\cdot z_xdx\\
&\ls&\alpha\|z_x\|^2+O(1)(\|f_1\|^2+\|f_2\|^2)\\
&\ls&\alpha\|z_x\|^2+O(\delta_T+\delta)^2
(\|z_{xx}\|^2+\|z_x\|^2)+O(\delta^2)(1+t)^ {-\frac{3}{2}},
\end{eqnarray*}
where $\alpha>0$ is a constant such that
$\alpha+O(\delta_T+\delta)^2\ls \frac{1}{10}\inf\limits_{x\in
\Real}p'(W)$. Combining these estimates, we get
\begin{equation*}
\frac{1}{2}\frac{d}{dt}\|z\|^2+(\frac{\varepsilon^2}{2}
-O(\delta_T+\delta)^2)
\|z_{xx}\|^2+\int_{\Real}(p'(W)-\alpha-O(\delta_T+\delta)^2)z_x^2dx
\ls O(\delta^2)(1+t)^{-\frac{3}{2}},
\end{equation*}
Since $\delta+\delta_T\ll 1$, we have
$\displaystyle\frac{\varepsilon^2}{2}-O(\delta+\delta_T)=O(\varepsilon^2)$
and
\begin{equation*}
\frac{1}{2}\frac{d}{dt}\|z\|^2+O(\varepsilon^2)\|z_{xx}\|^2
+\int_{\Real}(p'(W)- \alpha-O(\delta_T+ \delta))z_x^2dx \ls
O(\delta^2)(1+t)^{-\frac{3}{2}}.
\end{equation*}
Integrating the above inequality with respect to the time from 0 to
$t$, we get
$$
\|z(t)\|^2+\int_0^t\|z_{xx}(s)\|^2ds+\int_0^t\int_{\Real}
p'(W)z_x^2~dxds\ls O(\delta_0+\delta)^2.
$$
This, together with the fact $p'(W)>0$ for $W>0$, gives \ef{3.7}.
\end{proof}
\begin{lemma}{\label{lemma 3.5}}
Under the assumption \ef{3.1}, it holds for the local in-time
solution $z$
\bma
&\|z_x(t)\|^2+\displaystyle\int_0^t\|z_{xx}(s)\|^2ds+
\displaystyle\int_0^t \|z_{xxx}(s)\|^2ds
\ls O(\delta+\delta_0)^2,\label{3.8}\\
&(1+t)\|z_x(t)\|^2+\displaystyle\int_0^t(1+s)
(\|z_{xx}(s)\|^2+\|z_{xxx}(s)\|^2)ds\ls
O(\delta+\delta_0)^2\label{3.9} \ema for $0\ls t\ls T$ provided that
$\delta_T+\delta$ is small enough.
\end{lemma}
\begin{proof} Differentiating the equation \ef{2.5} with respect to $x$ and
taking the $L^2$-inner product of the resulting equation with $z_x$,
we get in view of the integration by parts \bma
&\int_{\Real}(f_1+f_2)_{xx}z_xdx\nnm\\
 =&\int_{\Real}
[({z_t})_x-((p'(W){z_x})_x)_x+\frac{\varepsilon^2}{2}({z_{xxxx})_x}]
\cdot z_x
dx\nnm\\
=&\frac{1}{2}\frac{d}{dt}\|z_x\|^2+\frac{\varepsilon^2}{2}
\|z_{xxx}\|^2+\int_{\Real} p'(W)z_{xx}^2dx+l_1,\label{3.10} \ema
where
 \be
 |l_1|=|\int_\Real p''(W)W_xz_xz_{xx}dx|\ls O(1)\|W_x\|^2_{L^{\infty}}
 \|z_x\|^2+\alpha \|z_{xx}\|^2,\label{3.11}
 \ee
with the same $\alpha$ as in Lemma \ref{lemma 3.4}. By Lemma
\ref{lemma 3.4} and Cauchy's inequality, we get
\bma |\int_{\Real}(f_1+f_2)_{xx}z_xdx|=&|\int_{\Real}
(f_1+f_2)_{x}z_{xx}dx|\nnm\\
\ls&O(1)\|(f_1+f_2)_x\|^2+\alpha\|z_{xx}\|^2
\nnm\\
\ls& O(\delta_T+\delta) (\|z_{xx}\|^2+\|z_{xxx}\|^2)
+\alpha\|z_{xx}\|^2 
+O(\delta^2)(1+t)^{-\frac{5}{2}},\label{3.12} \ema where we have
used the assumption \ef{3.1a} and a similar analysis as in Lemma
\ref{lemma 3.3} with the help of Lemma \ref{lemma 3.2}. Combining
\ef{3.10}--\ef{3.12} and integrating the resulting inequality over
$[0,t]$, we have after simplifying
$$
\|z_x(t)\|^2+\int_0^t\|z_{xx}(s)\|^2ds+\int_0^t\|z_{xxx}(s)\|^2ds
\ls O(\delta_0+\delta)^2,
$$
which gives \ef{3.8}. To obtain \ef{3.9},
 differentiating the equation \ef{2.5} with respect to
$x$ and taking the $L^2$-inner product of the resulting equation
with $(1+s)z_x$, similarly with the former we get after the
integration by parts \bma
 \int_{\Real}(f_1+f_2)_{xx}z_x(1+t)dx
 =&\int_{\Real}
[({z_t})_x-((p'(W){z_x})_x)_x+\frac{\varepsilon^2}{2}({z_{xxxx})_x}]
\cdot z_x(1+t)
dx\nnm\\
=&\frac{1}{2}\frac{d}{dt}[(1+t)\|z_x\|^2]-\frac{1}{2}\|z_x\|^2\nnm\\
&+\frac{\varepsilon^2}{2}(1+t) \|z_{xxx}\|^2+\int_{\Real}
(1+t)p'(W)z_{xx}^2dx+(1+t)l_1,\label{3.14} \ema with the estimate
\bma
 |(1+t)l_1|=&|\int_\Real (1+t)p''(W)W_xz_xz_{xx}dx|\nnm\\
 \ls& O(1)(1+t)\|W_x\|^2_{L^{\infty}}
 \|z_x\|^2+\alpha(1+t) \|z_{xx}\|^2\nnm\\
 \ls&O(1)(\delta^2)\|z_x\|^2+\alpha(1+t) \|z_{xx}\|^2,\label{3.15}
 \ema
where we have used Lemma \ref{lemma 3.2} for
$\|W_x\|^2_{L^{\infty}}\ls c\delta^2(1+t)^{-1}$. We also have
estimates \bma \int_{\Real}(1+t)(f_1+f_2)_{xx}z_xdx=&|\int_{\Real}
(1+t)(f_1+f_2)_{x}z_{xx}dx|\nnm\\
\ls& O(\delta_T+\delta) (1+t)(\|z_{xx}\|^2+\|z_{xxx}\|^2)
+\alpha(1+t)\|z_{xx}\|^2\nnm\\
&+O(\delta^2)(1+t)^{-\frac{3}{2}}.\label{3.16} \ema
Combining
\ef{3.14}--\ef{3.16} and integrating
 the resulting inequality over $[0,t_1]$ will give us after simplification
$$(1+t_1)\|z_x(t_1)\|^2+\displaystyle\int_0^{t_1}(1+t)
(\|z_{xx}(t)\|^2+\|z_{xxx}(t)\|^2)dt\ls O(\delta+\delta_0)^2,$$
where we have applied the derived
 estimates in Lemma \ref{lemma 3.4} to get that
$$\int_0^{t_1}\|z_x\|^2dt\ls O(\delta+\delta_0)^2$$
for $t_1\in [0,T]$. This gives \ef{3.9}.
\end{proof}

 Based on Lemma \ref{lemma 3.4} and Lemma \ref{lemma 3.5},
 we
can perform the estimates of the second and the third order
derivatives of $z$ in the same procedure and we have

\begin{lemma}{\label{lemma 3.6}}
 Under the assumption \eqref{3.1},
it holds for the local in-time solution $z$
\bma
&(1+t)^2\|z_{xx}(t)\|^2
+\displaystyle\int_0^t(1+s)^2(\|z_{xxx}(s)\|^2 +\|z_{xxxx}(s)\|^2)ds
\ls O(\delta+\delta_0)^2,\nnm\\
&(1+t)^3\|z_{xxx}(t)\|^2
+\displaystyle\int_0^t(1+s)^3(\|z_{xxxx}(s)\|^2
+\|z_{xxxxx}(s)\|^2)ds \ls O(\delta+\delta_0)^2\nnm \ema for $0\ls
t\ls T$ provided that $\delta_T+\delta$ is small enough.
\end{lemma}
\begin{proof} We give the sketch of the proof. Performing
$\int_0^{t_1}\int_\Real (1+t)^2(2.5)_{xx}z_{xx}dxdt$, we can get as
in Lemma \ref{lemma 3.5}
 \bma
&(1+t_1)^2\|z_{xx}(t)\|^2
+\displaystyle\int_0^{t_1}(1+t)^2(\|z_{xxx}(s)\|^2
+\|z_{xxxx}(s)\|^2)dt\nnm\\
 \ls& \int_0^{t_1}(1+t)^2\|z_x\|^2_{L^{\infty}}\|z_{xx}\|^2dt
 +\int_0^{t_1}(1+t)^2\|r_4\|^2dt,\label{3.17}
 \ema
with $r_4$ defined in Lemma \ref{lemma 3.3} satisfying
$$\|r_4\|^2\ls c\delta^2(1+t)^{-\frac{7}{2}}.$$
By \ef{3.1a} we also have $\|z_x\|^2_{L^{\infty}}\ls
c\delta_T^2(1+t)^{-1}$. Thus, by \ef{3.17} and Lemma \ref{lemma 3.5}
we have \bma (1+t_1)^2\|z_{xx}(t)\|^2
+\displaystyle\int_0^{t_1}(1+t)^2(\|z_{xxx}(s)\|^2
+\|z_{xxxx}(s)\|^2)dt
 \ls O(\delta+\delta_0)^2.\label{3.18}
 \ema
Similarly, by performing $\int_0^{t_1}\int_\Real
(1+t)^3(2.5)_{xxx}z_{xxx}dxdt$ and with the help of all the derived
a-priori estimates we can obtain
$$(1+t)^3\|z_{xxx}(t)\|^2
+\displaystyle\int_0^t(1+s)^3(\|z_{xxxx}(s)\|^2
+\|z_{xxxxx}(s)\|^2)ds \ls O(\delta+\delta_0)^2.$$
\end{proof}

\begin{proof}[\underline{The proof of Proposition~\ref{proposition 2.1}}]
 The Lemmas \ref{lemma 3.4}--\ref{lemma 3.6} show that the local
solution satisfies the uniform bounds for short time
($\delta\ll1,\delta_0\ll1$) when the initial perturbation is small
enough. By using continuous methods, we can extend the local
solution to be a global one, which also satisfies Lemmas \ref{lemma
3.4}--\ref{lemma 3.6} for any time. Then the existence of the global
solution to the IVP \ef{2.5}-\ef{2.6} is proven. The Theorem
\ref{theorem 1.1} is a direct corollary of Proposition
\ref{proposition 2.1}.
\end{proof}

\subsection{The bipolar case}
\label{bipolar}
In this subsection, we prove Theorem~\ref{theorem 1.2} about the
IVP~\ef{b.1}-\ef{b.1a} for the bipolar case. We also shall transform
the primary equations in order to study the existence and in
particular its large time behavior of
 global solutions of the IVP~\ef{b.1}-\ef{b.1a}.

Denote
\begin{align}\label{b.2}
    z^{\bh}(x,t)=\int_{-\infty}^x \left(\rho_{\bh}(y,t)-
    W\left(\frac{y+x_0}{\sqrt{t+1}}\right)\right)dy, \quad i=a,b,
\end{align}
which implies
\begin{align}\label{b.3}
    z^{\bh}_{x}=\rho_{\bh}(x,t)-W\left(\frac{x+x_0}{\sqrt{t+1}}\right), \quad i=a,b.
\end{align}

Similar to the unipolar case, assuming $(\rho_{\bh})_{xx}\to 0$ as
$|x|\to\infty$, we can rewrite the equations \eqref{b.1} as
\begin{align}\label{b.4}
    \left\{\begin{aligned}
    &z^{\bh}_t-(p'(W)z^{\bh}_{x})_x+\frac{\eps^2}{2}z^{\bh}_{xxxx}
    +(-1)^{i+1}(z^{\bh}_{x}+W)E=(f_{\bh,1}+f_{\bh,2})_x, \quad i=a,b,\\
    &E_t-(p'(W)E_x)_x+\frac{\eps^2}{2}E_{xxxx}+2WE
    =(f_{a,1}-f_{b,1}+f_{a,2}-f_{b,2})_x
    -(z^{a}_x+z^{b}_x)E,
    \end{aligned}\right.
\end{align}
with initial data
 \be
 z^{\bh}(x,0)=z^{\bh}_{0}(x), \quad
 z^{\bh}_{0}(\pm\infty)=0, \quad  i=a,b,   \label{b.4a}
 \ee
where
\begin{align*}
    E=&\int_{-\infty}^x (\rho_{a}(y,t)-\rho_{b}(y,t))dy=z^{a}-z^{b},\\
    f_{\bh,1}=&\frac{\eps^2}{2}\frac{(W_x+z^{\bh}_{xx})^2}{W+z^{\bh}_{x}}
    -\frac{\eps^2}{2}W_{xx}, \quad i=a,b,\\
    f_{\bh,2}=&p(z^{\bh}_{x}+W)-p'(W)z^{\bh}_{x}-p(W), \quad i=a,b,
\end{align*}
and we recall we make use of the symbols $(-1)^a=-1$ and $(-1)^b=1$
for the simplicity of statements.

We have the following results:

\begin{proposition}{\label{proposition 2.2}}
Let $p'(\rho)>0$ for $\rho>0$ and $\delta=|\rho_+-\rho_-|\ll1$.
Assume that $\|z^a_0(x)\|_{H^3(\Real)}+\|z^b_0(x)\|_{H^3(\Real)} \ls
\delta_0$ with $\delta_0>0$ sufficiently small, then there is a
unique global strong solution $(z^a,z^b,E)$ to the
IVP~\eqref{b.4}-\eqref{b.4a} satisfying
\begin{gather*}
(z^a,z^b)\in (L^{\infty}([0,\infty);H^3(\Real))\cap
L^2([0,\infty);H^5(\Real)))^2,
\\
\|\partial_x^k(z^a,z^b)(\cdot,t)\|_{L^2(\Real)} \ls
C(\delta,\delta_0)(1+t)^{-\frac{k}{2}},\ \ \ k=0,1,2,3,
\\
\|\partial_x^k(z^a,z^b)(\cdot,t)\|_{L^{\infty}(\Real)}\ls
C(\delta,\delta_0)(1+t)^{-\frac{1+2k}{4}},\ \ \ k=0,1,2,
\\
\|E(t)\|_{L^2(\Real)} \ls C(\delta,\delta_0)e^{-\beta t},
\end{gather*}
where $C(\delta,\delta_0)>0$ and $\beta $ are some positive
constants depending only on $\delta$ and $\delta_0$.
\end{proposition}

The proof of Proposition~\ref{proposition 2.2} can be made in the
similar fashion as Proposition~\ref{proposition 2.1}. Here, we only
show how to deal with the electric field and show its exponential
decay.  Let us assume that it holds for local in-time solutions that
\begin{align}\label{b.5}
\delta_T=&\max\limits_{k=0,1,2,3\atop i=a,b}\sup\limits_{0\ls t\ls
T}(1+t)^{\frac{k}{2}}\|\partial_x^kz^{\bh}\|\ll 1,\quad \text{for }
T>0.
\end{align}

Similar to Lemma~\ref{lemma 3.3}, we have the following properties
for $f_{\bh,1}$ and $f_{\bh,2}$, whose proof is very similar to that
of Lemma~\ref{lemma 3.3}, we omit details.

\begin{lemma}\label{lem.b.1}
Under the assumption \eqref{b.5},  we have for $i=a,b$ and
$j=0,1,2$, that
\begin{gather*}
    \partial_x^jf_{\bh,1}=
    O(\delta_T+\delta)\partial_x^{j+2}z^{\bh}+O(\delta)r_{j+2},\quad
    \partial_x^jf_{\bh,2}= O(\delta_T)\partial_x^{j+1}z^{\bh},\\
    \partial_x^j(f_{a,1}-f_{b,1})= O(\delta_T+\delta)
    (\partial_x^{j+1}E+\partial_x^{j+2}E),\quad
    \partial_x^j(f_{a,2}-f_{b,2})= O(\delta_T)\partial_x^{j+1}E,
\end{gather*}
where the function $r_k(x,t)$ is the same one as in Lemma~\ref{lemma
3.3}.
\end{lemma}

We also have the bipolar version of Lemmas~\ref{lemma
3.2}--\ref{lemma 3.6} as similar ways as in the unipolar case.
Indeed, we can obtain the following a-priori estimates.

\begin{lemma}\label{lem.b.2}
Under the assumption \eqref{b.5}, it holds for the strong solutions
$(z^{a},z^{b},E)$ that
\begin{align}\label{b.6}
    &\norm{\partial_x^k(z^{a},z^{b})(t)}^2
    +\sum_{j=1}^2\int_0^t\norm{\partial_x^{k+j}
    (z^{a},z^{b})(s)}^2ds +\int_0^t\norm{\partial_x^kE(s)}^2ds\ls
    O(\delta+\delta_0)^2,\\
    &(1+t)^k\norm{\partial_x^k(z^{a},z^{b})(t)}^2
    +\sum_{j=1}^2\int_0^t(1+s)^k\norm{\partial_x^{k+j}
    (z^{a},z^{b})(s)}^2ds\ls O(\delta+\delta_0)^2,\label{b.8}
\intertext{for $k=0,1,2,3,$ and}
    &\norm{\partial_x^kE(t)}\ls O(\delta+\delta_0)e^{-\beta_k t},
    \; k=0,1,2,\;\text{ for some constants } \beta_0>\beta_1,\beta_2>0, \label{b.7}
\end{align}
for $t\in[0,T]$, provided that $\delta_T+\delta$ is small enough.
\end{lemma}

\begin{proof}
We only give a sketch of the proof here since it is very similar to
the unipolar case. Taking the $k$th order derivative of the first
equation in \eqref{b.4} with respect to the spatial variable $x$,
and taking the $L^2$ inner product of the resulting equation with
$\partial_x^k z^{\bh}$, we get for $k=0,1,2,3$, with the help of
\eqref{b.5} and Lemma~\ref{lemma 3.2} and \ref{lem.b.1}, that
\begin{align*}
    &\sum_{i=a,b}\frac{1}{2}\frac{d}{dt}\norm{\partial_x^kz^{\bh}}^2
    +\sum_{i=a,b}\int_\Real
    p'(W)(\partial_x^{k+1}z^{\bh})^2dx+\frac{\eps^2}{2}\sum_{i=a,b}
    \norm{\partial_x^{k+2}
    z^{\bh}}^2+\int_\Real W(\partial_x^k E)^2dx\\
    =&(-1)^{\delta_{0k}}\sum_{i=a,b}\int_\Real
    \partial_x^{(k-1)(1-\delta_{0k})}(f_{\bh,1}+f_{\bh,2})
    \partial_x^{k+2-\delta_{0k}}z^{\bh}dx\\
    &-(1-\delta_{0k})\sum_{i=a,b}\sum_{j=0}^{k-1}C_k^j\int_\Real
    \partial_x^{k-j-1}(p''(W)W_x)\partial_x^{j+1}z^{\bh}\partial_x^{k+1}z^{\bh}dx\\
    &+\frac{1}{2}(-1)^k(\delta_{0k}+\delta_{1k})
    \int_\Real
    (z^{a}-z^{b})E\partial_x^{k+k+1}z^{a}dx\\
    &-(\delta_{2k}+\delta_{3k})
    \int_\Real\partial_x^{k-2}
    ((z^{a}_x-z^{b}_x)E)\partial_x^{k+2}z^{a}dx\\
    &-(-1)^{\delta_{1k}}\int_\Real\partial_x^{\delta_{3k}}
    (z^{b}_xE)\partial_x^{2k-\delta_{3k}}Edx
    -\sum_{j=0}^{k-1}C_k^j \int_\Real\partial_x^{k-j}W\partial_x^j E\partial_x^kEdx\\
    \ls &(\alpha_1+O(\delta^2))\sum_{i=a,b}\norm{\partial_x^{k+1+\min(k,1)}z^{\bh}}^2
    +O(\delta_T+\delta)^2\sum_{i=a,b}\norm{\partial_x^{2+\max(k-1,0)}z^{\bh}}^2\\
    &+O(\delta_T^2)\norm{\partial_x^{1+\max(k-1,0)}z^{\bh}}^2+\alpha_2\sum_{i=a,b}\norm{\partial_x^{k+1}z^{\bh}}^2\\
    &    +O(\delta^2)\frac{k}{\max(k,1)}\sum_{i=a,b}\sum_{j=0}^{k-1}C_k^j
       \norm{\partial_x^{j+1}z^{\bh}}^2
       +(\delta_{0k}+\delta_{1k})O(\delta_T^2)(\norm{E}^2
       +\norm{\partial_x^{2k+1}z^{a}})\\
       &+(\delta_{2k}+\delta_{3k})\Big[O(\delta_T^2)(\norm{E}^2
       +\delta_{3k}\norm{E_x}^2)
       +\alpha_3\norm{\partial_x^{k+2}z^{a}}^2\Big]\\
       &+\alpha_4\norm{\partial_x^kE}^2
       +O(\delta^2)\sum_{j=0}^{k-1}C_k^j\norm{\partial_x^jE}^2,
\end{align*}
where $\delta_{ij}=1$ if $i=j$ and $\delta_{ij}=0$ if $i\neq j$,
$\alpha_j$'s are small positive constants such that for sufficiently
small $\delta_T$, $\delta$ and $T$ ,
$p'(w)-\alpha_1-\alpha_2-O(\delta_T+\delta)^2>0$, $\frac{\eps^2}{2}-
\alpha_3-O(\delta_T^2)>0$ and $W-\alpha_4-O(\delta_T^2)>0$. Thus, it
is easy to obtain the desired result \eqref{b.6} by considering, in
turn, every $k=0,1,2,3$, respectively.

Now, we prove the second result \eqref{b.7}. Taking the $k$th  order
derivative of the second equation in \eqref{b.4} with respect to the
spatial variable $x$, and taking the $L^2$ inner product of the
resulting equation with $\partial_x^k E$, we have for $k=0,1,2$, in
view of \eqref{b.5} and Lemmas~\ref{lemma 3.2} and \ref{lem.b.1},
that
\begin{align*}
    &\frac{1}{2}\frac{d}{dt}\norm{\partial_x^kE(t)}^2
    +\delta_{0k}\int_\Real
    p'(W) E_x^2dx+\frac{\eps^2}{2}
    \norm{\partial_x^{k+2}E}^2
    +2\int_\Real W(\partial_x^kE)^2dx\\
     =&\delta_{1k}\int_\Real
    p'(W)E_xE_{xxx}dx+\delta_{2k}\int_\Real
    p''(W)W_xE_xE_{xxxx}dx
    -2(1-\delta_{0k})(-1)^k\int_\Real W
    E\partial_x^{2k}Edx\\
    &+2\delta_{1k}\int_\Real W(\partial_x^kE)^2dx-2\delta_{2k}
    \int_\Real WE_x\partial_x^3 Edx+2\delta_{0k}
    \int_\Real(z^{a}+z^{b})EE_xdx\\
    &-\delta_{1k}\int_\Real(z^{a}+z^{b})(E_xE_{xx}+EE_{xxx})dx+\delta_{2k}\int_\Real(z^{a}+z^{b})(3E_{xx}E_{xxx}+E_xE_{xxxx})dx\\
    &-(-1)^k(1-\delta_{2k})\int_\Real(f_{a,1}-f_{b,1}
    +f_{a,2}-f_{b,2})
    \partial_x^{2k+1}Edx\\
    &+\delta_{2k}\int_\Real(f_{a,1}-f_{b,1}
    +f_{a,2}-f_{b,2})_x
    \partial_x^{4}Edx\\
    \ls &\delta_{1k}[O(1)\norm{E_x}^2+\alpha_5\norm{E_{xxx}}^2]
    +\delta_{2k}[O(\delta^2)\norm{E_x}^2+\alpha_6\norm{E_{xxxx}}^2]\\
    &+(1-\delta_{0k})[O(\delta^2)\norm{E}^2+\alpha_7\norm{\partial_x^{2k}E}^2]
    +\delta_{2k}O(\delta^2)[\norm{E_x}^2+\norm{E_{xxx}}^2]\\
    &+\delta_{0k}[O(\delta_T^2)\norm{E}^2+\alpha_8\norm{E_x}^2]
    +\delta_{1k}O(\delta_T^2)[\norm{E_x}^2+\norm{E_{xx}}^2]
    +\delta_{1k}[O(\delta_T^2)\norm{E}^2+\alpha_9\norm{E_{xxx}}^2]\\
    &+\delta_{2k}O(\delta_T^2)[\norm{E_{xx}}^2+\norm{E_{xxx}}^2]
    +\delta_{2k}[O(\delta_T^2)\norm{E_x}^2+\alpha_{10}\norm{E_{xxxx}}^2]\\
    &+\delta_{0k}[O(\delta+\delta_T)\norm{E_x}^2+O(\delta+\delta_T)^2
    \norm{E_x}^2+\alpha_{11}\norm{E_{xx}}^2]\\
    &+\delta_{1k}
    [O(\delta+\delta_T)^2(\norm{E_x}^2+\norm{E_{xx}^2})+\alpha_{12}\norm{E_{xxx}}^2]\\
    &+\delta_{2k}[O(\delta+\delta_T)^2(\norm{E_{xx}}^2+\norm{E_{xxx}^2})
    +\alpha_{12}\norm{E_{xxxx}}^2],
\end{align*}
for some sufficiently small constants $\alpha_j$'s.  Thus, by
considering every $k=0,1,2$ in turn, there exist
$\beta_0>\beta_1,\beta_2>0$ for $\delta$ and $T$ small enough such
that
\begin{align*}
    \frac{1}{2}\frac{d}{dt}\norm{\partial_x^kE(t)}^2
    +\beta_k\norm{\partial_x^kE}^2
    \ls (1-\delta_{0k})O(\delta_T+\delta)\norm{E}^2, \quad k=0,1,2.
\end{align*}
From Gronwall inequality, we can obtain
\begin{align*}
    \norm{\partial_x^kE(t)}\ls O(\delta_0+\delta)e^{-\beta_kt}, \quad
    k=0,1,2.
\end{align*}

Finally, let us prove the last result \eqref{b.8}. Taking the $k$th
order derivative of the first equation in \eqref{b.4} with respect
to the spatial variable $x$, and taking the $L^2$ inner product of
the resulting equation with $(1+t)^k\partial_x^k z^{\bh}$, we get
for $k=1,2,3$, with the help of \eqref{b.5}, \eqref{b.6},
\eqref{b.7} and Lemma~\ref{lemma 3.2} and \ref{lem.b.1}, that
\begin{align*}
    &\frac{d}{dt}\left[\frac{(1+t)^k}{2}
    \sum_{i=a,b}\norm{\partial_x^k z^{\bh}}^2\right]
    +(1+t)^k\sum_{i=a,b}\int_\Real
    p'(W)(\partial_x^{k+1}z^{\bh})^2dx\\
    &\qquad\qquad+\frac{\eps^2}{2}(1+t)^k\sum_{i=a,b}
    \norm{\partial_x^{k+2}z^{\bh}}^2\\
    =&\frac{k}{2}(1+t)^{k-1}\sum_{i=a,b}\norm{\partial_x^k
    z^{\bh}}^2-(1+t)^k\sum_{i=a,b}\sum_{j=0}^{k-1}C_k^j\int_\Real\partial_x^{k-j}
    p'(W)\partial_x^{j+1}z^{\bh}\partial_x^{k+1}z^{\bh}dx \\
    &-2(1+t)^k\int_\Real W(\partial_x^k E)^2dx-(1+t)^k\sum_{i=a,b}\sum_{j=0}^{k-1}C_k^j \int_\Real
    \partial_x^{k-j}(z_x^{\bh}+W)\partial_x^jE\partial_x^kEdx\\
    &-(1+t)^k\sum_{i=a,b}\int_\Real z_x^{\bh}(\partial_x^k E)^2 dx
    +(1+t)^k\sum_{i=a,b}\int_\Real
    \partial_x^{k-1}(f_1^{\bh}+f_2^{\bh})\partial_x^{k+2}z^{\bh}dx\\
    \ls &\frac{k}{2}(1+t)^{k-1}\sum_{i=a,b}\norm{\partial_x^k
    z^{\bh}}^2
    +\delta_{1k}(1+t)^k\sum_{i=a,b} \left[\alpha_{14}\norm{z_{xxx}^{\bh}}^2
    +O(1)(p'(W))^2\norm{z_x^{\bh}}^2\right]\\
    &+\delta_{2k}(1+t)^k\sum_{i=a,b}\left[O(\delta^2)\norm{z_{xx}^{\bh}}^2
    +\alpha_{15}\norm{z_{xxxx}^{\bh}}^2+O(1)\norm{z_{xxx}^{\bh}}^2\right]\\
    &+\delta_{3k}(1+t)^k\sum_{i=a,b}\left[O(\delta^2)\left(\norm{z_{xx}^{\bh}}^2
    +\norm{z_{xxx}^{\bh}}^2\right)+O(1)\norm{z_{xxxx}^{\bh}}^2
    +\alpha_{16}\norm{z_{xxxxx}^{\bh}}^2\right]\\
    &+O(\delta+\delta_T)O(\delta+\delta_0)^2,
\end{align*}
which implies the desired result as the way as the previous.
\end{proof}

\bigskip

\begin{proof}[\underline{The proof of Theorem~\ref{theorem 1.2} and Proposition~\ref{proposition 2.2}}]
Since the proof is very similar to the unipolar case in view of
Lemma~\ref{lem.b.2}, we omit details.
\end{proof}

\vspace{0.5cm}

\noindent\textbf{Acknowledgments}:\  The authors would like to thank
the referees for helpful comments and suggestions on the
presentation of the manuscript.

The authors acknowledge the partial supports of the National Science
Foundation of China (No.10431060, No.10571102, No.10601061), the Key
Research Project on Science and Technology of the Ministry of
Education of China (No.104072), Beijing Nova program, the Institute
of Mathematics and Interdisciplinary Science at CNU, the NCET
support of the Ministry of Education of China, the Huo Ying Dong
Foundation No.111033, and the Scientific Research Startup Special
Foundation for the Winner of the Award for Excellent Doctoral
Dissertation and the Prize of President Scholarship of Chinese
Academy of Sciences.

\end{document}